# Truthful Mechanisms via Greedy Iterative Packing[*]

Chandra Chekuri[†]    Iftah Gamzu[‡]

**Abstract**

An important research thread in algorithmic game theory studies the design of efficient truthful mechanisms that approximate the optimal social welfare. A fundamental question is whether an $\alpha$-approximation algorithm translates into an $\alpha$-approximate truthful mechanism. It is well-known that plugging an $\alpha$-approximation algorithm into the VCG technique may not yield a truthful mechanism. Thus, it is natural to investigate properties of approximation algorithms that enable their use in truthful mechanisms.

The main contribution of this paper is to identify a useful and natural property of approximation algorithms, which we call loser-independence; this property is applicable in the single-minded and single-parameter settings. Intuitively, a loser-independent algorithm does not change its outcome when the bid of a losing agent increases, unless that agent becomes a winner. We demonstrate that loser-independent algorithms can be employed as sub-procedures in a greedy iterative packing approach while preserving monotonicity. A greedy iterative approach provides a good approximation in the context of maximizing a non-decreasing submodular function subject to independence constraints. Our framework gives rise to truthful approximation mechanisms for various problems. Notably, some problems arise in online mechanism design.

[*]An extended abstract of this paper appeared in *Proceedings of the 12th International Workshop on Approximation Algorithms for Combinatorial Optimization Problems*, 2009.

[†]Department of Computer Science, University of Illinois, Urbana, IL 61801. Partially supported by NSF grants CCF-0728782 and CNS-0721899. Email: `chekuri@cs.illinois.edu`.

[‡]Blavatnik School of Computer Science, Tel-Aviv University, Tel-Aviv 69978, Israel. Supported by the Binational Science Foundation, by the Israel Science Foundation, by the European Commission under the Integrated Project QAP funded by the IST directorate as Contract Number 015848, and by a European Research Council (ERC) Starting Grant. Email: `iftgam@tau.ac.il`.

# 1  Introduction

Algorithmic aspects of mechanism design have become an important area of research in recent years. A central research theme focuses on the design of efficient mechanisms for algorithmic problems in strategic settings. These mechanisms must take into account both standard computational efficiency considerations and strategic behavior of the participants. The latter goal commonly correlates with the development of truthful mechanisms, namely, mechanisms that are robust against manipulation by the participants. The primary technique of mechanism design, i.e., VCG mechanisms [42, 18, 26], is known to be truthful for optimizing social welfare. Unfortunately, implementing VCG is computationally intractable in many (even simple) settings of interest since the underlying optimization problem that needs to be solved is NP-Hard. An important research thread in algorithmic game theory, starting with the work of Nisan and Ronen [38], focussed on designing efficient truthful mechanisms that approximate the optimal social welfare. A fundamental question is whether an $\alpha$-approximation algorithm translates into an $\alpha$-approximate truthful mechanism. It is well-known that plugging an $\alpha$-approximation algorithm into the VCG mechanism may not yield a truthful mechanism [32, 39]. Thus, it is natural to investigate properties of approximation algorithms that enable their use in truthful mechanisms.

The problem of combinatorial auctions has gained the status of the paradigmatic problem in the field of algorithmic mechanism design. For a detailed overview, see [11]. In the context of single-minded agents, Lehmann, O'Callaghan and Shoham [32] established that an approximation algorithm can support a truthful mechanism if it satisfies a monotonicity property. Consequently, monotone approximation algorithms and techniques have been developed for various combinatorial optimization problems that underlie special cases of combinatorial auctions such as multi-unit auctions [35, 12]. One interesting set of techniques, devised by Mu'alem and Nisan [35], enables one to combine approximation algorithms while preserving monotonicity. In particular, they identified a special case of monotonicity, which they name bitonicity, and demonstrated that bitonic algorithms may be combined via the "max" operation.

## 1.1  Our results

The main contribution of this paper is to identify a useful and natural property of approximation algorithms, which we name *loser-independence*. Intuitively, a loser-independent algorithm does not change its outcome when the bid of a losing agent increases, unless that agent becomes a winner. We demonstrate that loser-independent algorithms can be employed as sub-procedures in a greedy iterative packing approach while preserving monotonicity. A greedy iterative approach provides good approximation in the context of maximizing a non-decreasing submodular function subject to independence constraints such as matroid constraints [36, 23, 25]. There are various interesting problems that can be cast as special instances of this family (see, e.g., [14, 43]). We note that our loser-independence property is somewhat orthogonal to the notion of *composability* presented by Aggarwal and Hartline [1]. Intuitively, a composable algorithm does not change its outcome when the bid of a winning agent varies above its critical winning bid. Moreover, combining our property with the composability property yields, in the current setting, the *stability* condition suggested by Dobzinski and Sundararajan [20]. This condition states that if the bid of an agent changes but its allocation stays the same then the allocations to all other agents also do not change.

Our framework gives rise to efficient truthful approximation mechanisms for several problems. Notably, some of these problems arise in online mechanism design. We view the framework and the identification of the loser-independence property as the key contribution, and hence, we focus on those rather than the improvements for specific problems. We illustrate the applicability of the framework by briefly outlining two representative results that we derive.

**An offline setting.** A truthful $(2+\epsilon)$-approximate mechanism for the multiple knapsack problem among single-minded agents. This result improves and generalizes a 6-approximation mechanism for a special case of multiple knapsack among single-parameter agents [4]. In addition, we show that an almost identical mechanism attains an approximation ratio of $2 + \epsilon$ for the generalized assignment problem among single-parameter agents. This is the first non-trivial approximate truthful mechanism for this problem when the number of knapsacks is part of the input; a monotone PTAS exists for this problem when the number of knapsacks is a fixed constant [12].

**An online setting.** A truthful 2-competitive mechanism for the online problem of dynamic auction with expiring items. This mechanism is essentially identical to the mechanism devised by Hajiaghayi et al. [28]. Furthermore, we achieve a truthful $(2+\epsilon)$-competitive mechanism for the generalization of the problem in which the underlying auction in each time-slot is a multi-item auction among single-minded agents rather than a single-item auction.

## 1.2  Related work

It is widely known that many common techniques that are broadly used by approximation algorithms cannot be used in a strategic setting since they violate certain monotonicity properties which are imperative for truthfulness. Correspondingly, recent years have seen an ever-growing line of work addressing the development of monotone algorithmic alternatives. Mu'alem and Nisan [35] seem to have been the first to pay attention to this issue. They presented sufficient conditions for composing monotone algorithms via two basic operators, namely "max" and "if-then-else". Briest, Krysta and Vöcking [12] devised a general approach to transform a pseudo-polynomial algorithm into a monotone FPTAS, and demonstrated that primal-dual greedy algorithms may be used to derive truthful mechanisms. Lavi and Swamy [31] designed a general technique to convert approximation algorithms in packing domains to randomized approximation mechanisms that are truthful in expectation. Babaioff, Lavi and Pavlov [8] presented a method that translates any given algorithm to a truthful mechanism in single parameter domains. However, their method degrades the performance guarantee of the resulting mechanism by a factor of $O(\log \rho)$, where $\rho$ denotes the ratio between the largest and smallest valuations. Recently, Azar and Gamzu [4] presented a monotone partition framework for approximating packing integer programs.

Focusing on the previously-mentioned representative problems from a purely algorithmic point of view, the multiple knapsack problem is known to admit a PTAS by the work of Chekuri and Khanna [15], while the generalized assignment problem is known to be approximable within a factor that is slightly better than $e/(e-1) \approx 1.582$ by the work of Feige and Vondrák [22]. The dynamic auction with expiring items problem is equivalent to online scheduling of unit-length jobs on a single machine to maximize weighted throughput. The best known deterministic online algorithm for this problem has a competitive ratio of about 1.828 [21] (see also [33]), while it is known that no deterministic online algorithm can achieve a competitive ratio better than $\phi \approx 1.618$ [27, 17, 2]. Turning to the randomized setting, the best online algorithm attains a ratio of $e/(e-1)$ [9, 16], while it is known that no online algorithm can attain a ratio better than 1.25 [17]. This problem can be solved optimally in the offline setting.

## 2  The General Setting

In this section, we study the truthfulness properties of an iterative packing approach for a general class of maximizing assignment problems with packing constraints. We illustrate our ideas by restricting attention to the separable assignment problem [24]. In Section 4, we discuss our approach in the context of maximizing a non-decreasing submodular function subject to independence constraints. Note that the separable assignment problem is an instance of maximizing a non-decreasing submodular function over a partition matroid [24, 14].

An instance of the single-parameter variant of the *separable assignment problem* consists of a collection $B$ of $m$ bins and a set $U$ of $n$ items. Each bin $j \in B$ has a separable independence system $I_j \subseteq 2^U$, representing the subsets of items that may be packed in that bin[1]. Each item $i \in U$ has a positive value $v_i$, which is gained by assigning the item to one of the bins. The objective is to find a maximum value subset of items $S \subseteq U$, along with an assignment of these items to the bins, so that all the items in $S$ can be simultaneously placed in their designated bins, while preserving the constraints induced by the independence systems. In particular, the set of items assigned to bin $j$, namely, $S_j$, must satisfy $S_j \in I_j$. In the game theoretic version of this problem there are $n$ strategic single parameter agents, each of which controls an item, and may be untruthful about its value.

We consider the following *iterative packing approach* for approximately solving the mentioned problem: assume the existence of an $\alpha$-approximation oracle for the single bin sub-problem, and build a solution by iteratively packing each of the bins (without backtracking) using the oracle. More precisely, the single bin sub-problem corresponding to bin $j$ is to find a maximum value subset $S_j$ that satisfies $S_j \in I_j$, and the iterative packing approach utilizes the approximation oracle to generate a packing $S_1 \subseteq U$ for the first bin, then it is used to generate a packing $S_2 \subseteq U \setminus S_1$ for the second bin, and so on. In what follows, we investigate the truthfulness properties of the iterative packing approach. Specifically, we focus on the approximation oracle, and establish a sufficient condition which guarantees that the iterative approach will lead to a monotone algorithm, and hence, a truthful mechanism.

### 2.1  Preliminaries

We introduce some notation and terminology that will be used throughout the paper, and describe a characterization that links monotone algorithms with truthful mechanisms. The reader is encouraged to refer to [37, 11] for a more comprehensive overview of the underlying concepts.

We will mainly concentrate on two types of agents: single-parameter and single-minded. *Single-parameter* agents have private data that consists of a single number, namely, their value. *Single-minded* agents [32, 12] have private data which consists of a pair $(o, v)$, where $o$ is an object that the agent is interested in and $v$ is the valuation of the agent for attaining $o$. Remark that the interpretation of the object $o$ depends on the problem at hand. For example, an object may represent a bandwidth demand of an agent (as in network routing), and it might stand for a set of items that an agent wants (as in combinatorial auctions). The valuation function of an agent whose data is $(o, v)$ is a step function with respect to $o$. Specifically, if the agent obtains the object $o$ or any object that extends it then its valuation is $v$; otherwise, its valuation is 0. We use the notation $\tilde{o} \preceq o$ to indicate that object $o$ *extends* object $\tilde{o}$. Again, the interpretation of the term extension depends on the problem at hand. For instance, if the object $\tilde{o}$ represents bandwidth demand then the object $o$ extends it if it represents a higher bandwidth demand, and if the object $\tilde{o}$ stands for a set of items then the object $o$ extends it if it stands for a superset of the items.

---

[1] An *independence system* $I \subseteq 2^U$ is a family of subsets that is downward closed, that is, $A \in I$ and $B \subseteq A$ implies that $B \in I$. Note that the packing constraints are implicit from the independence systems which guarantee that if some subset of items is feasible for a bin then any subset of it is also feasible.

We now present the notion of monotonicity for single-minded agents, and then turn to describe a characterization that reduces the goal of designing truthful mechanisms to that of designing monotone algorithms. Note that similar definitions can be made for single-parameter agents by refining the monotonicity property and the characterization theorem. Specifically, both of them need to be defined only with respect to the value of every agent, and the objects-related terms need to be cast off. We say that an agent is a *selected* if it is assigned the object $o$ or any object that extends it.

**Definition 2.1.** An algorithm $\mathcal{A}$ is said to be *monotone* with respect to the bid of an agent if it satisfies the following property: if algorithm $\mathcal{A}$ selects the agent when its bid is $(o, v)$ then it selects the agent when its bid is $(\tilde{o}, \tilde{v})$, where $\tilde{o} \preceq o$ and $\tilde{v} \geq v$, and the bids of all the other agents are fixed.

**Theorem 2.2.** ([12]) *If algorithm $\mathcal{A}$ is monotone with respect to the bid of every agent then there exists a corresponding truthful mechanism which can be efficiently computed using algorithm $\mathcal{A}$.*

### 2.2 A motivating example

Let us consider the single-parameter variant of the multiple knapsack problem. This problem is a special case of the separable assignment problem, where the single bin sub-problem is the *knapsack problem*. Specifically, an instance of the *multiple knapsack problem* consists of a collection $B$ of $m$ bins, and a set $U$ of $n$ items. Each bin $j \in B$ has a capacity $W_j$, and each item $i \in U$ is characterized by a pair $(w_i, v_i)$, where $w_i$ is the size of the item and $v_i$ is its positive value. The goal is to select a maximum value subset of items $S \subseteq U$, along with an assignment of these items to the bins, so that all the items in $S$ can be simultaneously placed in their designated bins while preserving the capacities of the bins. Note that the single bin sub-problem that corresponds to bin $j$ is to find a maximum value subset of items whose overall size does not exceed $W_j$. Also notice that the independence system of bin $j$ is $I_j = \{S \in 2^U : \sum_{i \in S} w_i \leq W_j\}$.

---
**Algorithm 1** MaxGreedy

   **Input:** A set of items $U$, and the capacity of the bin $W$
   **Output:** A set of items $S$ to be assigned to the bin
1: $U_1 \leftarrow U$, $U_2 \leftarrow U$, $S_1 \leftarrow \emptyset$, $S_2 \leftarrow \emptyset$
2: **while** $U_1 \neq \emptyset$ **do**
3:    **remove** the item $i$ that has a maximum value from $U_1$
4:    **if** $\sum_{\ell \in S_1} w_\ell + w_i \leq W$ **then add** $i$ to $S_1$
5: **end while**
6: **while** $U_2 \neq \emptyset$ **do**
7:    **remove** the item $i$ that has a maximum profit density from $U_2$
8:    **if** $\sum_{\ell \in S_2} w_\ell + w_i \leq W$ **then add** $i$ to $S_2$
9: **end while**
10: **return** the maximum value allocation between $S_1$ and $S_2$

---

We focus on algorithm MaxGreedy, formally described above, which approximately solves the knapsack problem. This algorithm initially computes two assignments: one based on a greedy approach with respect to the values of the items, and another based on a greedy approach with respect to the *profit density* ratio of the items, that is, a value to size ratio. Then, it returns the assignment having maximum value. This algorithm was considered by Mu'alem and Nisan [35], who proved that it is monotone with respect to the value, and that it achieves 2-approximation. Due to its monotone properties, it may seem natural to use this algorithm as the single bin approximation oracle in the iterative approach attending to the multiple knapsack problem. Unfortunately, as the following theorem states, the resulting iterative algorithm fails to be monotone.

**Theorem 2.3.** *The iterative packing approach that employs algorithm MaxGreedy as the single bin approximation oracle is not monotone.*

**Proof.** Suppose we are given an input instance that consists of two unit-capacity bins, and the following set of items $U = \{(1/2, 1+\epsilon), (1/2, 1+\epsilon), (3/4, 3/2), (1/4, 1/2), (1, 2-\epsilon), (1, 2-\epsilon)\}$, where $\epsilon < 1/6$. Notice that the first two items have profit density of $2 + 2\epsilon$, the next two items have profit density of 2, and the last two items have profit density $2 - \epsilon$. Also note that the last two items have maximum value. One can validate that the iterative packing approach returns an allocation in which the first bin consists of the pair of items whose profit density is $2 + 2\epsilon$ and the second bin consists of the pair of items whose profit density is 2. Particularly, the single bin oracle returns the allocation generated by the greedy approach with resect to profit density ratio for both bins. This solution is schematically described in Figure 1(a). Now, suppose that the value of the selected item $(1/4, 1/2)$ increases to $1/2 + \epsilon$, while the values of all other items are fixed. Notice that the profit density ratio of that item increases to $2 + 4\epsilon$. If the iterative packing approach was monotone, this item would still be allocated. Unluckily, this is not the case. One can verify that the iterative packing approach returns an allocation in which the two maximum value items are packed in the bins, as schematically illustrated in Figure 1(b). ∎

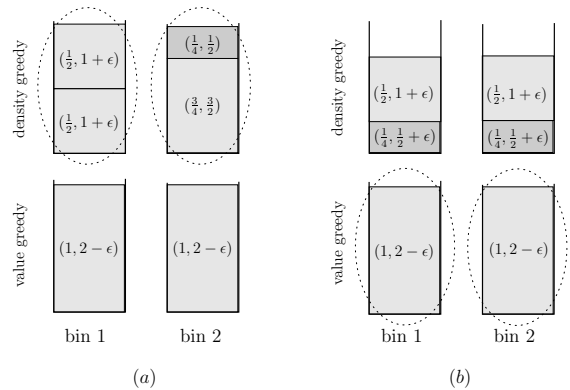

Figure 1: The outcome of the iterative packing approach with respect to the input instances described in the proof. The $(w, v)$ pair inscribed in each block corresponds to the size and value of the item.

It is worth noting that algorithm MaxGreedy is not only monotone, but also bitonic with respect to the value [35]. Informally, an algorithm is *bitonic* if its outcome value as a function of the value of any single agent $i$ has the pattern that it does not increase as long as agent $i$ is not selected, and it does not decrease as long as agent $i$ is selected. This implies that both monotonicity and bitonicity of the single bin approximation oracle are not sufficient to ensure the monotonicity of the corresponding iterative packing approach.

## 2.3 A sufficient condition

In the following, we establish a sufficient condition for the single bin approximation oracle. This condition guarantees that the iterative packing approach, which employs the oracle as the single bin sub-procedure, will satisfy monotonicity. We present the condition for single-minded agents, and an analogous condition for single-parameter agents can be derived in a similar manner.

We briefly motivate the sufficient condition by using the algorithm and the multiple knapsack instance described in the previous subsection. Recall that monotonicity guarantees that if a selected agent improves its bid then it continues to be selected. In particular, the monotonicity of algorithm MaxGreedy implies that if an agent that is selected for the first bin increases its value then it continues to be selected for that bin. However, the key difficulty appears when we consider an agent that was selected in a later bin, say the second one. In this case, when that agent increases its value then (from the perspective of the first bin) it is like a non-selected agent increases its value. Consequently, no guarantees can be made with respect to the assignment generated for the first bin. As a result, the agent that increased its value may compete against a different set of agents for the second bin, and may not be selected. One way to deal with this difficulty is to restrict the set of algorithms that may be employed by the iterative packing approach to those that are loser-independent, as formally defined below. Intuitively, a loser-independent algorithm does not disturb an assignment when the value of a losing agent increases, unless that agent becomes a winner. Note that this requirement is satisfied by any optimal (1-approximation) algorithm.

**Definition 2.4.** An algorithm $\mathcal{A}$ is said to be *loser-independent* with respect to the bid of an agent if it satisfies the following property: if algorithm $\mathcal{A}$ generates the solution $a$ in which the agent is not selected when its bid is $(o, v)$ then algorithm $\mathcal{A}$ either generates the same solution $a$ or selects the agent when its bid is $(\tilde{o}, \tilde{v})$, where $\tilde{o} \preceq o$ and $\tilde{v} \geq v$, and the bids of all the other agents are fixed.

**Theorem 2.5.** *If algorithm $\mathcal{A}$ is loser-independent and monotone with respect to the bid of every agent then the iterative packing approach, which employs it as the single bin oracle, is monotone with respect to the bid of every agent.*

**Proof.** Consider an agent $i$ that is selected to the $j$-th bin by the algorithm when its bid is $(o, v)$. Now, suppose that $i$ improves its bid to $(\tilde{o}, \tilde{v})$ such that $\tilde{o} \preceq o$ and $\tilde{v} \geq v$, and the bids of all the other agents are fixed. For the sake of monotonicity, we need to prove that the iterative packing approach selects $i$ to some bin in the latter case. If $i$ is selected in one of the first $j - 1$ packing iterations of the algorithm, that is, the iterations that pack the first $j - 1$ bins, then we are clearly done. Otherwise, let us focus on the $j$-th iteration. Notice that during the first $j - 1$ packing iterations, the same set of agents are selected. This follows from the loser-independent property of algorithm $\mathcal{A}$, which guarantees that none of the assignments of the first $j - 1$ bins was disturbed. Accordingly, the same set of agents remain to compete for bin $j$, which implies, in conjunction with the monotonicity property of algorithm $\mathcal{A}$, that agent $i$ must be selected in the $j$-th iteration. ∎

The above characterization can be easily extended for a collection of algorithms $\mathcal{A} = \{A_1, \ldots, A_m\}$ that are loser-independent and monotone. Specifically, one can prove that if each algorithm $A_j$ is loser-independent and monotone with respect to the bid of every agent then the iterative packing approach, which employs $A_j$ as the oracle of bin $j$, is monotone with respect to the bid of every agent. We defer the details to the full version of this paper.

## 2.4 Applications

The domain of problems for which the mentioned characterization is useful is broad. Essentially, one may take any single-minded or single-parameter version of a social welfare maximization packing problem $\pi$, and design a new "multiple" variant of this problem via the separable assignment problem paradigm. In what follows, we demonstrate the applicability of the characterization by utilizing it in the context of several well-known or highly-motivated problems. Note that the characterization reduces the task of designing a monotone algorithm for the "multiple" variant of $\pi$ to that of designing a loser-independent and monotone algorithm for $\pi$.

We begin by pointing out two known approximation properties of the iterative packing approach. We will utilize these properties when analyzing the approximation ratio of the iterative approach for problems under consideration. The first property states that given an $\alpha$-approximation oracle for the single bin sub-problem of the separable assignment problem, the iterative packing approach has an approximation ratio of at most $\alpha + 1$ [23, 13, 25]. The second property pertains to the special case of the separable assignment problem in which all the separable independence systems are identical, that is, $I_1 = \cdots = I_m$. In this case, it is known that given an $\alpha$-approximation oracle for the single bin sub-problem, the iterative packing approach achieves approximation ratio of at most $e^{1/\alpha}/(e^{1/\alpha} - 1)$ [36, 25]. Note that these properties are additional motivation for our use of the iterative packing approach as they show that the approximation ratio of the iterative approach for the "multiple" variant problem degrades by constants with respect to that of the single bin oracle. For purpose of self-containment, formal proofs of these properties appear in Appendix A.1.

**The multiple knapsack and generalized assignment problems.** In the following, we consider the single-minded variant of the multiple knapsack problem, and the single-parameter variant of the generalized assignment problem. Both problems are special cases of the separable assignment problem, where the single bin sub-problem is the knapsack problem. The single-minded variant of multiple knapsack generalizes the single-parameter variant presented in Subsection 2.2 by allowing each agent $i$ to be dishonest about the pair $(w_i, v_i)$, that is, it may by untruthful about both the size and the value of the corresponding item. The single-parameter variant of the *generalized assignment problem* extends the single-parameter variant of multiple knapsack by characterizing each item $i$ with a pair $(w_i, v_i)$, where $w_i$ is a *vector* of length $m$ which represents the size that item $i$ occupies in each of the bins, and $v_i$ is its positive value. Note that the private data of agent $i$ consists only of the value $v_i$, while the vector $w_i$ is public knowledge.

We demonstrate the utility of the new condition via these problems. We begin by designing a relatively simple 2-approximate algorithm for the knapsack problem that may be utilized as the single bin oracle. It is instructive to measure our algorithm against algorithm MaxGreedy as our algorithm is loser-independent. Algorithm HalfGreedy, formally described below, begins by computing two assignments: one which consists only of the maximum value item, and another that is based on a greedy approach with respect to the profit density ratio. Note that profit density greedy approach in our algorithm has three key differences from the profit density greedy approach applied by algorithm MaxGreedy. The first is that it only considers *small* items, namely, items whose size is no more than half the capacity of the bin; the second is that it stops adding items to the assignment once their overall size is at least half of the capacity of the bin; and the third is that it defines the value of the assignment, marked by $V_2$, to be the overall value of the items that fill exactly half of the capacity of the bin (unless the overall size of all small items is less than that). Particularly, in the former case, only a portion of the value of the last item included in the assignment is taken into account. This portion corresponds to the portion of the size of the item contained before the half-way mark of the bin. Note that $V_2$ is a lower bound on the overall value of the items in the profit density greedy assignment. Then, the algorithm returns the assignment corresponding to a greater assignment value.[2]

---

**Algorithm 2** HalfGreedy

**Input:** A set of items $U$, and the capacity of the bin $W$
**Output:** A set of items $S$ to be assigned to the bin

1: $U_1 \leftarrow U$, $U_2 \leftarrow \{i \in U : w_i \leq W/2\}$, $S_1 \leftarrow \emptyset$, $S_2 \leftarrow \emptyset$
2: **let** $S_1$ be the singleton set that consists of the maximum value item of $U_1$
3: **let** $V_1$ be the value of the single item in $S_1$
4: **while** $\left(\sum_{\ell \in S_2} w_\ell < W/2 \text{ and } U_2 \neq \emptyset\right)$ **do**
5:     **remove** the item $i$ that has a maximum profit density from $U_2$
6:     **add** $i$ to $S_2$
7: **end while**
8: **let** $i_1, \ldots, i_k$ be the items selected to $S_2$ according to their inspection order.
9: **let** $V_2 = \sum_{\ell=1}^{k-1} v_{i_\ell} + v_{i_k}/w_{i_k} \cdot \min\{w_{i_k}, W/2 - \sum_{\ell=1}^{k-1} w_{i_\ell}\}$
10: **if** $V_1 \geq V_2$ **then return** $S_1$ **else return** $S_2$

---

For purposes of analyzing the monotonicity and loser-independent properties of the algorithm, we consider two input instances that are identical with the exception that in the first instance the bid of agent $i$ is $(w, v)$ and in the second instance its bid is $(\tilde{w}, \tilde{v})$ such that $\tilde{w} \leq w$ and $\tilde{v} \geq v$. We denote the assignments and corresponding assignment values that algorithm HalfGreedy generates for the first instance by $S_1, S_2$ and $V_1, V_2$, respectively. Similarly, we mark the assignments and corresponding values generated for the second instance by $\tilde{S}_1, \tilde{S}_2$ and $\tilde{V}_1, \tilde{V}_2$. We begin by proving the following lemma, which plays an instrumental role in establishing the mentioned properties.

**Lemma 2.6.** *Consider an index $k \in \{1, 2\}$. The corresponding assignment values satisfy $\tilde{V}_k \geq V_k$. Furthermore, if $\tilde{V}_k > V_k$ then $i \in \tilde{S}_k$, and if $\tilde{V}_k = V_k$ then either $i \in \tilde{S}_k$ or $\tilde{S}_k = S_k$.*

**Proof.** Recall that $V_1$ is the value of the maximum value item. It is clear that this value can only increase when any item increases its value. Thus, $\tilde{V}_1 \geq V_1$. Moreover, if $\tilde{V}_1 > V_1$ then $i$ must be the maximum value item, and hence, $i \in \tilde{S}_1$. Lastly, consider the case $\tilde{V}_1 = V_1$. We assume that possible ties are broken according to some fixed ordering on the set of items. This implies that the assignment could not have changed, unless $i$ increased its value to $V_1$, become part of the group of maximum value items, and was selected.

Let us turn to the profit density greedy approach. Recall that this approach packs the most profitable small items in terms of value to size, and sets the assignment value to be the overall value of the items that fill exactly half of the capacity of the bin. The only exception is if the overall size of all small items is less than half of the capacity of the bin, and then, the assignment value is the overall value of all the small items. We begin by considering the simple case in which the overall size of all the small items in the second instance, denoted by $\tilde{U}_2$, is less than half the capacity of the bin. In this case, $\tilde{S}_2 = \tilde{U}_2$. Notice that the set of small items in the first instance, namely, $U_2$, satisfies $U_2 \subseteq \tilde{U}_2$. This follows by recalling that the only difference between both instances is that the bid of $i$ changed such that its size may have decreased. Hence, $S_2 \subseteq U_2 \subseteq \tilde{U}_2 = \tilde{S}_2$, which implies that $\tilde{V}_2 \geq V_2$. Let us now consider the case that the overall size of all the small items in the second instance is at least half of the capacity of the bin. The claim that $\tilde{V}_2 \geq V_2$ follows by observing that if we place the items of $\tilde{S}_2$ in the first half of the bin according to an non-increasing profit density ratio then any point in this packing has a profit density ratio that is at least as large as the profit density ratio of that point in the corresponding packing according to $S_2$. Here, the profit density ratio of a point is the profit density ratio of the item the contains it. For the purpose of proving the two additional properties, notice that if $i \notin \tilde{S}_2$ then $i \notin S_2$. This results since the algorithm is greedy with respect to profit density ratio, and $i$ has better profit density ratio (and smaller size) in the second instance. In turn, this implies that if $i \notin \tilde{S}_2$ then the assignments and the corresponding assignment values generated for both instances should be identical as $i$ plays no role in any of them. Consequently, if $\tilde{V}_2 = V_2$ and $i \notin \tilde{S}_2$ then $\tilde{S}_2 = S_2$. In addition, if $\tilde{V}_2 > V_2$ then $i \in \tilde{S}_2$. ∎

---

[2]There are high-level similarities between algorithm HalfGreedy and algorithm AK of [1]; however, the properties for which the algorithms designed for are quite different. We thank Tim Roughgarden for pointing out [1].

We are now ready to prove the main properties of our algorithm.

**Theorem 2.7.** *Algorithm* HalfGreedy *achieves* 2*-approximation and maintains monotonicity and loser-independence with respect to the bid of every agent.*

**Proof.** We begin by proving the approximation guarantee of the algorithm. Let $S^* = \{i_1^*, \ldots i_\ell^*\}$ be the set of items packed in the optimal solution ordered according to a non-increasing size, and let $\text{OPT} = \sum_{j=1}^{\ell} v_{i_j^*}$ be the overall value of the items in $S^*$. We consider two cases, depending whether $S^*$ contains a big item or not, and demonstrate that in both cases the approximation ratio of the algorithm is 2. An item is *big* if its size is more than half the capacity of the bin.

*Case I: $S^*$ consists of a big item.* Notice that $S^*$ can consist of only one big item, that is, $i_1^*$. In addition, the overall size of all the other (necessarily small) items is less than half the capacity of the bin. This implies that $V_2 \geq \sum_{j=2}^{\ell} v_{i_j^*}$. Moreover, it is clear that $V_1 \geq v_{i_1^*}$. Hence,

$$\max(V_1, V_2) \geq \frac{1}{2}(V_1 + V_2) \geq \frac{1}{2} \sum_{j=1}^{\ell} v_{i_j^*} = \frac{1}{2}\text{OPT} .$$

*Case II: $S^*$ only consists of small items.* Recall that the profit density greedy approach generates an assignment in which either all small items are packed or at least half of the bin is packed with the most profitable small items in terms of value to size. Thus, it follows that

$$\max(V_1, V_2) \geq V_2 \geq \frac{1}{2} \sum_{j=1}^{\ell} v_{i_j^*} = \frac{1}{2}\text{OPT} .$$

We turn to prove that the algorithm is monotone with respect to the bid of every agent. For this purpose, we utilize the previously mentioned input instances, and argue that if agent $i$ is selected by the algorithm with respect to the first instance then $i$ must be selected with respect to the latter instance. In what follows, we establish the claim for the case that $V_1 \geq V_2$, noting that the complementary case can be proved by using nearly identical arguments. Note that $i \in S_1$. We consider two cases, depending on the assignment values attained by the algorithm for the second instance, and demonstrate that in both of them $i$ is selected:

- *Case I: $\tilde{V}_1 \geq \tilde{V}_2$.* Notice that Lemma 2.6 guarantees that in any case either $i \in \tilde{S}_1$ or $\tilde{S}_1 = S_1$. Nevertheless, since $i \in S_1$, it follows that $i \in \tilde{S}_1$.

- *Case II: $\tilde{V}_2 > \tilde{V}_1$.* Observe that $\tilde{V}_2 > \tilde{V}_1 \geq V_1 \geq V_2$, where the second inequality results from Lemma 2.6. This implies, in conjunction with Lemma 2.6, that $i \in \tilde{S}_2$.

We now establish that the algorithm is loser-independent with respect to the bid of every agent. Similarly to before, we utilize the aforementioned input instances, and claim that if agent $i$ is not selected by the algorithm with respect to the first instance then it is either selected with respect to the latter instance or the algorithm generates the same assignment. In the following, we prove the claim for the case that $V_2 > V_1$, noting that the complementary case can be proved by using nearly identical arguments. Note that $i \notin S_2$. We consider two cases, depending on the assignment values obtained by the algorithm for the second instance:

- Case I: $\tilde{V}_1 \geq \tilde{V}_2$. Notice that $\tilde{V}_1 \geq \tilde{V}_2 \geq V_2 > V_1$, where the second inequality results from Lemma 2.6. This implies, in conjunction with Lemma 2.6, that $i \in \tilde{S}_1$.

- Case II: $\tilde{V}_2 > \tilde{V}_1$. Lemma 2.6 guarantees that in any case either $i \in \tilde{S}_2$ or $\tilde{S}_2 = S_2$. ∎

Theorem 2.7, Theorem 2.5 and the previously mentioned approximation properties, which are formally stated in Lemma A.1 and Lemma A.2, imply the following corollaries.

**Corollary 2.8.** *There is a truthful $3$-approximation mechanism for the generalized assignment problem among single-parameter agents.*

**Corollary 2.9.** *There is a truthful $3$-approximation mechanism for multiple knapsack among single-minded agents. This mechanism attains an approximation ratio of $e^{1/2}/(e^{1/2} - 1) \approx 2.541$ when bin capacities are identical.*

As we are interested in better performance guarantees, we turn to study the monotone FPTAS for the knapsack problem, developed by Briest, Krysta and Vöcking [12]. We prove that this algorithm is loser-independent, and thus, may be employed by an iterative packing approach. A formal description of this algorithm, referred to as algorithm MonotoneFPTAS, is given below.[3] Algorithm MonotoneFPTAS enumerates over a collection of $O(\log(n/\epsilon))$ algorithms. Each of these algorithms scales the values of the items in a different fashion, and then applies the well-known pseudo-polynomial time algorithm for knapsack, referred to as PseudoPack, which computes an optimal assignment with respect to the scaled values. Finally, the algorithm returns the maximum value assignment with respect to the de-scaled values. Note that if several assignments have the same maximum value then it is assumed that this tie is broken according to the index $k$.

**Theorem 2.10.** *Algorithm MonotoneFPTAS is a fully polynomial time approximation scheme that maintains monotonicity and loser-independence with respect to the bid of every agent.*

**Proof.** The approximation guarantee and the monotonicity of the algorithm are established in [12]. Hence, we are left to prove that the algorithm is loser-independent. Essentially, this is accomplished using a similar line of argumentation as exhibited in the proofs of Lemma 2.6 and Theorem 2.7.

Consider two input instances that are identical with the exception that in the first instance the bid of agent $i$ is $(w, v)$ and in the second instance its bid is $(\tilde{w}, \tilde{v})$, where $\tilde{w} \leq w$ and $\tilde{v} \geq v$. Let $K$ be the collection of scaling indices used by the algorithm with respect to the first instance, namely, the algorithm employs the parameterized sub-procedure $\mathsf{Scale}_k$, for every $k \in K$. Moreover, let $S_k$ and $V_k$ be the assignment and corresponding value that the algorithm generates by employing $\mathsf{Scale}_k$ with respect to the first instance. Specifically, $S_k$ is the assignment that $\mathsf{Scale}_k$ generates and $V_k$ is

---

[3]Note that MonotoneFPTAS differs from the algorithm presented in [12] with respect to the collection of the scaling algorithms it employs. Particularly, our algorithm considers a superset of scaling algorithms. This modification does not alter any of the properties proved in [12], but simplifies the analysis of the loser-independence property.

---
**Algorithm 3** MonotoneFPTAS
---
**Input:** A set of items $(v, w)$, where $v$ and $w$ are value and weight vectors of the items
**Output:** A set of items $s$ to be assigned, where $s$ is a $\{0,1\}$-assignment vector of the items

1: $s \leftarrow \langle 0, \ldots, 0 \rangle$, best $\leftarrow 0$
2: Let $V$ be the maximum value of an item
3: **for** $k \leftarrow \lceil \log(\frac{nV}{\epsilon}) \rceil, \ldots, \lceil \log(V) \rceil - \log(\frac{n}{1-\epsilon}) - 1$ **do**
4: $\quad (\tilde{s}, \tilde{v}) \leftarrow \mathsf{Scale}_k(v, w)$
5: $\quad$ **if** $\tilde{s} \cdot \tilde{v} >$ best **then**
6: $\quad\quad s \leftarrow \tilde{s}$, best $\leftarrow \tilde{s} \cdot \tilde{v}$
7: $\quad$ **end if**
8: **end for**
9: **return** $s$
10: **procedure** $\mathsf{Scale}_k(v, w)$
11: $\quad \alpha \leftarrow \frac{n}{\epsilon \cdot 2^k}$
12: $\quad$ let $\bar{v}$ be the truncated values vector in which each $\bar{v}_i = \min\{v_i, 2^k\}$
13: $\quad$ let $v'$ be the scaled values vector in which each $v'_i = \lfloor \alpha_k \cdot \bar{v}_i \rfloor$
14: $\quad$ let $\tilde{v}$ be the de-scaled values vector in which each $\tilde{v}_i = v'_i / \alpha_k$
15: $\quad$ **return** $(\mathsf{PseudoPack}(v', w), \tilde{v})$
16: **end procedure**
---

the value of this assignment according to the de-scaled values. In a similar manner, let $\tilde{S}_k$ and $\tilde{V}_k$ be the assignment and corresponding value attained by $\mathsf{Scale}_k$ with respect to the second instance.

Recall that $\mathsf{Scale}_k$ computes an optimal assignment with respect to the scaled values. Moreover, notice that the scaling process is independent of the values of the items and only depends on the parameter $k$. This implies, in conjunction with the fact that the bid of $i$ improves in the latter instance, that $\tilde{V}_k \geq V_k$. In addition, one can verify that the optimality of $\mathsf{Scale}_k$ and the properties of the scaling process ensure that if $\tilde{V}_k > V_k$ then $i \in \tilde{S}_k$, and if $\tilde{V}_k = V_k$ then either $i \in \tilde{S}_k$ or $\tilde{S}_k = S_k$. We are now ready to argue that if agent $i$ is not selected by the algorithm with respect to the first instance then it is either selected with respect to the second instance or the algorithm generates the same assignment. Suppose the algorithm returns the assignment $S_{k^*}$ when it is applied to the first instance. We consider three cases, depending on the assignment returned by the algorithm for the second instance:

- *Case I: the algorithm returns $\tilde{S}_{k^*}$.* The above-mentioned argument guarantees that in any case either $i \in \tilde{S}_{k^*}$ or $\tilde{S}_{k^*} = S_{k^*}$.

- *Case II: the algorithm returns $\tilde{S}_{k'}$ such that $k' \neq k^*$ and $k' \in K$.* We prove that $\tilde{V}_{k'} > V_{k'}$. Consequently, the above-mentioned argument ensures that $i \in \tilde{S}_{k'}$. Assume by contradiction that $\tilde{V}_{k'} = V_{k'}$. This implies that all the inequalities in the chain $\tilde{V}_{k'} \geq \tilde{V}_{k^*} \geq V_{k^*} \geq V_{k'}$ must be satisfied with equality. However, this contradicts the assumption that ties are broken in a consistent way.

- *Case III: the algorithm returns $\tilde{S}_{k'}$ where $k' \notin K$.* Notice that this case may only happen if the value of $i$ becomes strictly greater than the value of the maximum value item in the first instance, henceforth denoted by $V$, and in consequence, the collection of scaling algorithms changed. In particular, $k'$ must be greater than all the indices in $K$, that is, $k' > \lceil \log(\frac{nV}{\epsilon}) \rceil$.

Let us focus on the scaling of all the items except $i$ by $\mathsf{Scale}_{k'}$. Each such item has a value of at most $V$, and thus, it must be scaled to 0. This implies that the sole reason that $\tilde{S}_{k'}$ was returned by the algorithm is that $i \in \tilde{S}_{k'}$.  ∎

**Corollary 2.11.** *There is a truthful $(2+\epsilon)$-approximation mechanism for the generalized assignment problem among single-parameter agents.*

**Corollary 2.12.** *There is a truthful $(2+\epsilon)$-approximation mechanism for multiple knapsack among single-minded agents. This mechanism attains an approximation ratio of $e/(e-1) + \epsilon \approx 1.582$ when bin capacities are identical.*

**Additional applications.** In what follows, we briefly list several additional packing problems whose "multiple" variant can be solved by exploiting the characterization. In particular, we identify the corresponding loser-independent algorithms.

*Combinatorial auctions.* The *multi-unit* combinatorial auction problem (see, e.g., [3, 10]) is a natural generalization of the celebrated combinatorial auction problem in which each good has several copies. One may interpret the "multiple" variant of this problem as adding *group constraints* to the basic problem. Specifically, in this variant, each good is associated with a group, and goods from different groups cannot be used to form a bundle satisfying an agent. One can demonstrate that the algorithm for single-minded combinatorial auction [32], and the algorithms for the single-minded multi-unit version [12, 5] maintain loser-independence.

The *single value* combinatorial auction problem is a special case of the combinatorial auction problem in which the valuation of each agent is represented by a single value. Particularly, each (multi-minded) agent is interested in several different bundles, but obtains the same value from any non-zero outcome. It is clear that our characterization is not applicable since the agents are multi-minded. Still, if the agents are *known*, that is, all their data besides their values is publicly known, then one can establish that the characterization is still suitable. Essentially, this follows from the observation that truthfulness in known agents setting reduces to value monotonicity. Similarly to before, one may interpret the "multiple" variant of this problem as adding group constraints to the basic problem, and may prove that the algorithm for single value combinatorial auction among known multi-minded agents [7] maintains loser-independence.

*Advertisement space auctions.* The theme of selling advertisement space on a newspaper page can be modelled by packing convex figures in a plane. One may interpret the "multiple" variant of this problem as increasing the advertisement space to several pages, and may demonstrate that the algorithms presented in [6] maintain loser-independence.

*Network routing.* The task of routing in networks is commonly modelled using the unsplittable flow problem. One may interpret the "multiple" variant of this problem as adding wavelength constraints to the basic problem. These constraints prevent serving requests across different wavelengths. One can verify that the algorithms presented in [12, 5] maintain loser-independence.

## 3 The Online Setting

In this section, we extend our results for an online environment in which agents arrive and leave dynamically over time and there is uncertainty about the set of decisions to be made in the future. We illustrate our ideas by considering the online version of the separable assignment problem. In this variant, bins are aligned with discrete time slots, and items arrive and depart dynamically. In particular, an item is not known prior to its arrival and cannot be assigned after its departure. The goal is to generate a maximum value assignment of items to bins in an online fashion. Specifically, any assigned item must be packed in a bin that corresponds to a time slot between its arrival and departure times. In the game theoretic version of this problem, each agent controls an item, and may be untruthful about its value, arrival time, and departure time. In adherence with previous results in an online setting [41, 28], we assume *no early-arrival* and *no late-departure* misreports. That is, agents cannot report an arrival time earlier than their true arrival time or a departure time later than their true departure time. Note that Lavi and Nisan [30] considered the special case of separable assignment problem in which any bin can only accommodate a single item, and proved that it is impossible to attain bounded competitive ratio without restricting the misreports.

Our approach to solve this online variant is identical to before. Namely, we assume the existence of an $\alpha$-approximation oracle for the single bin sub-problem, and build a solution by iteratively employing it to generate an assignment for each of the bins. Notice that a single bin oracle optimizes with respect to a current state of agents and does not take into account the global system-wide view, and hence, if bins are considered according to their time order then the iterative approach constitute an online algorithm.

### 3.1 A sufficient condition

In what follows, we reformulate the sufficient condition for the single bin approximation oracle, exhibited in Subsection 2.3, for online environments. We begin by presenting revised definitions of single-minded agents and monotonicity for an online setting. Remark that the forthcoming definitions can be refined for single-parameter agents in a similar manner to before. Additionally, we encourage the reader to refer to [40] for a more detailed overview of online mechanisms.

The private data of single-minded agent in an online setting consists of a quadruple $(o, v, a, d)$, where $o$ is an object that the agent is interested in, $v$ is the valuation of the agent for attaining $o$, and $a$ and $d$ are the arrival and departure times of the agent, respectively.

**Definition 3.1.** *An online algorithm $\mathcal{A}$ is said to be* monotone *with respect to the bid of an agent if it satisfies the following property: if algorithm $\mathcal{A}$ selects the agent when its bid is $(o, v, a, d)$ then algorithm $\mathcal{A}$ selects the agent when its bid is $(\tilde{o}, \tilde{v}, \tilde{a}, \tilde{d})$, where $\tilde{o} \preceq o$, $\tilde{v} \geq v$, $\tilde{a} \leq a$ and $\tilde{d} \geq d$, and the bids of all the other agents are fixed.*

**Theorem 3.2.** ([40]) *If online algorithm $\mathcal{A}$ is monotone with respect to the bid of every agent then there exists a corresponding truthful mechanism which can be computed using algorithm $\mathcal{A}$.*

We are ready to prove that an online iterative packing approach, which employs a monotone and loser-independent oracle as the single bin sub-procedure, satisfies monotonicity. Note that the monotonicity and loser-independence of the oracle are with respect to the non-temporal part of the bid, that is, the object-value pair $(o, v)$.

**Theorem 3.3.** *If algorithm $\mathcal{A}$ is loser-independent and monotone with respect to the non-temporal bid of every agent then the online iterative packing approach, which employs it as the single bin oracle, is monotone with respect to the bid of every agent.*

**Proof.** Consider an agent $i$ that is selected to bin $j$ by the algorithm when its bid is $(o, v, a, d)$. Note that $j \in \{a, \ldots, d\}$. Now, suppose that $i$ improves its bid to $(\tilde{o}, \tilde{v}, \tilde{a}, \tilde{d})$ such that $\tilde{o} \preceq o$, $\tilde{v} \geq v$, $\tilde{a} \leq a$ and $\tilde{d} \geq d$, and the bids of all the other agents are fixed. Notice that $j \in \{\tilde{a}, \ldots, \tilde{d}\}$. For the sake of monotonicity, we need to prove that the online iterative packing approach selects $i$ to one of the bins $\{\tilde{a}, \ldots, \tilde{d}\}$ in the latter case. If $i$ is selected to one of the bins $\{\tilde{a}, \ldots, j-1\}$ then we are clearly done. Otherwise, let us focus on bin $j$. Notice that the same set of agents are selected to the first $j-1$ bins. This follows from the loser-independence property of algorithm $\mathcal{A}$ which guarantees that none of the assignments of the first $j-1$ bins was disturbed. It seems worthy to emphasize that if $i$ competes in time slots that it did not compete in the former instance, e.g., when $\tilde{a} < a$, one may interpret its prior non-attendance as if it was not selected. Accordingly, the same set of agents remain to compete for bin $j$, which implies, in conjunction with the monotonicity of algorithm $\mathcal{A}$, that agent $i$ must be selected to bin $j$. ∎

## 3.2 Applications

Similarly to the offline setting, the domain of problems for which the mentioned characterization is useful is broad. Basically, one may take any single-minded or single-parameter version of a social welfare maximization packing problem, and design an online variant of this problem via the online separable assignment problem paradigm. Several straightforward examples are the problems presented in Subsection 2.4. Note that the online iterative packing approach achieves a competitive ratio of at most $\alpha + 1$, assuming an $\alpha$-approximation oracle for the single bin sub-problem. This claim can be established by using nearly identical arguments to the ones presented in Lemma A.1.

An additional interesting application is the problem of *dynamic auction with expiring items*. This problem is a special case of the online separable assignment problem, where the single bin sub-problem is a *single-item auction*. An instance of this problem consists of a collection of unit-capacity bins, each associated with a distinct time-slot. An additional ingredient of the input is an online sequence of unit-size items, each of which is characterized by a triple $(v, a, d)$, where $v$ is its positive value, $a$ is its arrival time, and $d$ is its departure time. The objective is to generate a maximum value assignment of items to bins in an online fashion. In particular, this assignment should place at most one item in each bin, and each assigned item must be placed in a bin that corresponds to a time slot between its arrival and departure times. Focusing on the single bin sub-problem, one can notice that it admits a trivial optimal algorithm which places the most valuable item in a bin. As previously mentioned, any optimal algorithm is monotone and loser-independent. Hence, the characterization and the claimed approximation property imply the following corollary.

**Corollary 3.4.** *There is a truthful 2-competitive mechanism for dynamic auction with expiring items among single-parameter agents.*

Interestingly, this simple online iterative packing algorithm is identical to the algorithm presented by Hajiaghayi et al. [28], and it is best possible [28, 19]. Specifically, no deterministic truthful mechanism can obtain a competitive ratio better than 2. A natural generalization of this problem can be obtained by replacing the single bin sub-problem of single-item auction with *multi-item auction*. We refer to this problem as *dynamic auction with expiring multi-items*. It is well-known that the combinatorial optimization problem that underlie multi-unit auction among single-minded agents is the knapsack problem. In correspondence with previous results, we yield the following corollary.

**Corollary 3.5.** *There is a truthful $(2+\epsilon)$-competitive mechanism for dynamic auction with expiring multi-items among single-minded agents.*

## 4 Additional Applications via Submodular Function Maximization

As mentioned before, the greedy iterative approach provides good approximation in the context of maximizing a non-decreasing submodular function subject to independence constraints. More formally, let $f : 2^N \to \mathcal{R}^+$ be a non-decreasing submodular function on a finite ground set $N$, and let $(N, \mathcal{I})$ be an independence family. In other words, $\mathcal{I} \subseteq 2^N$ is a family of subsets that is downward closed, that is, $A \in \mathcal{I}$ and $B \subseteq A$ imply $B \in \mathcal{I}$. The optimization problem is then $\max_{S \in \mathcal{I}} f(S)$. Interesting independence families are matroids, intersection of a small number $k$ of matroids, and somewhat more general notions such as $k$-independence and $k$-extendible systems (see [29, 34, 14]). The greedy approach is then simple; start with an empty set, and incrementally build a solution by greedily adding an element that (approximately) improves the current solution the most while maintaining its independence. It is known that the greedy approach gives a $(k\alpha + 1)$-approximation for the above problem if there is an $\alpha$-approximation for picking the element that most improves the current solution [23] (see [25, 14] for recent and more easily available proofs).

When the underlying optimization problem of mechanism design, in particular the winner determination problem, can be cast as a special case of submodular function maximization subject to independence constraints, one may be able to use the greedy approach. In this case, if the (approximation) algorithm employed by the greedy incremental step is monotone and loser-independent then one can show that the overall greedy approach is monotone. We remark that the greedy approach here is somewhat different from the one presented in Section 2 for separable assignment problems; for the latter case, we employed a *local greedy* approach which considers the bins according to an *arbitrary* ordering and packs each bin with the approximate best solution. However, a global greedy approach would have considered all empty bins in each step and then pack the bin that most improves the solution. We note that the local greedy approach works for partition matroids [23], and is essential for the applications in online settings. Still, more general independence constraints requires the global greedy approach. As we remarked, loser-independence is still applicable. We give a concrete application to illustrate it.

Consider the multiple knapsack problem (or the generalized assignment problem), and suppose we add a constraint that at most $m' < m$ of the bins can be used in the packing. The resulting optimization problem becomes a submodular function maximization problem subject to a laminar matroid constraint (as observed in [14]). In this setting, the global greedy approach needs to pick in each step the best bin to pack by trying all remaining bins. One can easily extend Theorem 2.5 to this setting, and prove that if the single bin algorithm is monotone and loser-independent then the greedy approach is monotone. Details are deferred to the full version of this paper. We hope that additional applications to mechanism design problems will be found by using the above high-level approach.

**Acknowledgments:** The authors thank Yossi Azar, Jason Hartline, Tim Roughgarden and Jan Vondrák for useful discussions and comments on topics related to this paper.

## A  The General Setting

In this section, we present, for completeness, some details omitted from Section 2 of the paper. See [36, 15, 24, 13, 25] for results and proofs on greedy for multiple knapsack, generalized and separable assignment problems, and submodular function maximization subject to matroid and independence constraints.

### A.1  The performance guarantees of the iterative approach

In this subsection, we prove the claimed performance guarantee properties of the iterative packing approach. For ease of presentation, we introduce the following notation:

- Let $S^*$ denote the set of items packed in the optimal solution, and let $S$ be the set of items packed in the solution generated using the iterative packing approach. Furthermore, let $S = S_1 \cup \cdots \cup S_m$ be the partition of the items in $S$ into $m$ disjoint sets according to their assignment in the solution of the iterative approach. Specifically, $S_j$ consists of all the items of $S$ that were assigned to bin $j$ in that solution.

- Let $v(T) = \sum_{i \in T} v_i$ denote the overall value of items in $T$. Particularly, we let $\text{OPT} = v(S^*)$ and $\text{ALG} = v(S)$ be the value of the optimal solution and the solution generated using the iterative packing approach, respectively.

**Lemma A.1.** *The iterative packing approach achieves approximation ratio of at most $\alpha + 1$ given an $\alpha$-approximation oracle for the single bin sub-problem.*

**Proof.** Let $T = S^* \setminus S$ be the set of items that are packed in the optimal solution, but not packed by the solution of the iterative packing approach. Moreover, let $T = T_1 \cup \cdots \cup T_m$ be the partition of the items in $T$ into $m$ disjoint sets according to their assignment in the optimal algorithm. Now, if $v(T) \leq \alpha/(\alpha+1) \cdot \text{OPT}$ then we are clearly done since

$$\text{ALG} = v(S) \geq v(S^*) - v(T) \geq \frac{1}{\alpha+1}\text{OPT} \ .$$

Otherwise, notice that $v(S_\ell) \geq 1/\alpha \cdot v(T_\ell)$. This follows by the $\alpha$-approximation guarantee of the oracle, and the fact that $T_\ell$ was a feasible solution for the oracle when it considered bin $\ell$. Consequently, we get that

$$\text{ALG} = v(S) = \sum_{\ell=1}^m v(S_\ell) \geq \frac{1}{\alpha} \sum_{\ell=1}^m v(T_\ell) = \frac{1}{\alpha} v(T) > \frac{1}{\alpha+1}\text{OPT} \ ,$$

where the last inequality follows from the assumption that $v(T) > \alpha/(\alpha+1) \cdot \text{OPT}$. ∎

**Lemma A.2.** *The iterative packing approach has approximation ratio of at most $e^{1/\alpha}/(e^{1/\alpha} - 1)$ given an $\alpha$-approximation oracle for the single bin sub-problem, when the separable independence systems are identical.*

**Proof.** We begin by claiming that $v(S_j) \geq 1/(\alpha m) \cdot (OPT - v(\bigcup_{\ell=1}^{j-1} S_\ell))$, for every $j \in [m]$. For this purpose, notice that a simple counting argument proves that there is a set of items $T \subseteq S^* \setminus \bigcup_{\ell=1}^{j-1} S_\ell$ that is assigned to a single bin by the optimal algorithm and has $v(T) \geq 1/m \cdot (OPT - v(\bigcup_{\ell=1}^{j-1} S_\ell))$. Now, the claim follows by observing that $T$ was feasible for the $\alpha$-approximation oracle when it considered bin $j$ since all the separable independence systems are identical.

We turn to prove that $\text{OPT} - v(\bigcup_{\ell=1}^j S_\ell) \leq (1 - 1/(\alpha m))^j \cdot \text{OPT}$, for every $j \in \{0, \ldots, m\}$. This is achieved by induction on $j$. It is clear that this argument holds when $j = 0$. Now, notice that

$$\text{OPT} - v\left(\bigcup_{\ell=1}^j S_\ell\right) \leq \left(1 - \frac{1}{\alpha m}\right)\left(\text{OPT} - v\left(\bigcup_{\ell=1}^{j-1} S_\ell\right)\right) \leq \left(1 - \frac{1}{\alpha m}\right)^j \text{OPT} \ ,$$

where the first inequality follows from the above-mentioned claim, and the last inequality results from the induction hypothesis. In particular, this implies that $\text{OPT} - v(\bigcup_{\ell=1}^m S_\ell) \leq (1 - 1/(\alpha m))^m \cdot \text{OPT} \leq e^{-1/\alpha} \cdot \text{OPT}$. Consequently, we get that

$$\text{ALG} = v\left(\bigcup_{\ell=1}^m S_\ell\right) \geq \left(\frac{e^{1/\alpha} - 1}{e^{1/\alpha}}\right) \text{OPT} \ .$$

∎